\documentclass[twocolumn,amsmath,amssymb,aps,prl,lettersize]{revtex4-1}

\usepackage{soul}
\usepackage{xcolor}
\usepackage{amsmath}
\usepackage{txfonts}
\usepackage{amssymb}
\usepackage{dcolumn}

\newcommand{\rr}{\mathbf{r}}

\newcommand{\uu}{\mathbf{u}}
\newcommand{\atan}{\mathrm{atan}}

\usepackage{lipsum}
\usepackage{graphicx}
\usepackage{physics}
\usepackage{xcolor}
\usepackage{txfonts}
\usepackage[hypertexnames=false]{hyperref}
\usepackage[shortlabels]{enumitem}
\hypersetup{colorlinks=true,citecolor=blue,linkcolor=blue,urlcolor=blue}
\usepackage{dcolumn}
\usepackage{bm}
\newcommand{\fig}[1]{{Fig.~\ref{#1}}}
\newcommand{\eq}[1]{{Eq.~(\ref{#1})}}

\begin{document}

\title{Dynamical mechanisms of vortex pinning in superfluid thin films}

\author{Oliver R. Stockdale}
\affiliation{ARC Centre of Excellence in Future Low-Energy Electronics
Technologies, School of Mathematics and Physics, University of Queensland, St
Lucia, QLD 4072, Australia}

\author{Matthew T. Reeves}
\email{m.reeves@uq.edu.au}
\affiliation{ARC Centre of
Excellence in Future Low-Energy Electronics Technologies, School of Mathematics
and Physics, University of Queensland, St Lucia, QLD 4072, Australia}

\author{Matthew J. Davis} 
\affiliation{ARC Centre of
Excellence in Future Low-Energy Electronics Technologies, School of Mathematics
and Physics, University of Queensland, St Lucia, QLD 4072, Australia}

\date{\today}
\begin{abstract}
We characterize the mechanisms of vortex pinning in a superfluid thin film described by the two-dimensional Gross-Pitaevskii equation. We consider a vortex ``scattering experiment" whereby a single vortex in a superfluid flow interacts with a circular, uniform pinning potential. By an analogy with linear dielectrics, we
develop an analytical hydrodynamic approximation that predicts vortex trajectories, the vortex fixed point and the unpinning velocity. We then solve the Gross-Pitaevskii equation to validate this model, and build a phase portrait of vortex pinning. We identify two
    different dynamical pinning mechanisms marked by distinctive phonon emission
    signatures: one enabled by acoustic radiation and another mediated by vortex dipoles nucleated within the pin. Relative to obstacle size, we find that pinning potentials on the order of the healing length are more {effective} for vortex capture. Our results could be useful in mitigating the negative effects of drag due to vortices in superfluid channels, in analogy to maximising supercurrents in type-II superconductors.
\end{abstract}

\maketitle

\textit{Introduction}---The pinning of topological defects plays an important
role in many physical and biological systems including cardiac
muscle~~\cite{davidenko1992, pumir1999}, active matter~\cite{pazo2004}, and liquid crystals~\cite{campbell2014}.
In quantum fluids, such as superfluids and superconductors,
the defects are quantized vortices. Their pinning and unpinning from potential barriers is crucial in
determining the breakdown of dissipationless superflow~\cite{donev2001}. For example, vortex pinning is important for preventing flux creep in high-$T_c$
superconductors~\cite{nelson1993,blatter1994,kwok2016}, 
corrections to the Berezinskii–Kosterlitz–Thouless transition in thin-film
He-II~\cite{hedge1980,ambegaokar1980, adams1987}, and manipulating flows in atomtronic  devices~\cite{kuopanportti2010a,neely2013,samson2016}. Beyond these lab-based systems, sudden rotation `glitches' observed in neutron stars are hypothesized to involve an abrupt `avalanche' of vortices unpinning \emph{en masse} from the star's outer crust~\cite{anderson1975,epstein1988,warszawski2012}. Further, the recent achievement of
room-temperature superfluids~\cite{lerario2017} has stimulated interest in
harnessing superfluidity in future quantum technologies such as low-energy transistors~\cite{zasedatelev2019}. Similar to superconductors, a better understanding of vortex pinning may allow for enhanced superfluid critical currents.

While vortex nucleation is well-understood in terms
of Landau's critical velocity~\cite{landau1941,frisch1992,jackson2000,jung2020}, the subsequent mechanisms for vortex pinning
and unpinning still lack a complete theoretical description. For
finite temperature systems which can be modelled by e.g., 
Ginzburg-Landau equations~\cite{kwok2016,pazo2004} or two-fluid
models~\cite{schwarz1981}, a
vortex will be  gradually attracted to a (stable) dynamical fixed point through the dissipative action of the normal fluid component.  However, this cannot explain how
pinning occurs in a pure superfluid with minimal thermal friction, a regime in which experiments now routinely operate~\cite{bradley2008,niemetz2017,gauthier2019b,sachkou2019}. This poses the question:~What are the microscopic
mechanisms for vortex pinning in a pure superfluid system? 

\begin{figure}
    \centering
    \includegraphics[width= 0.9\columnwidth]{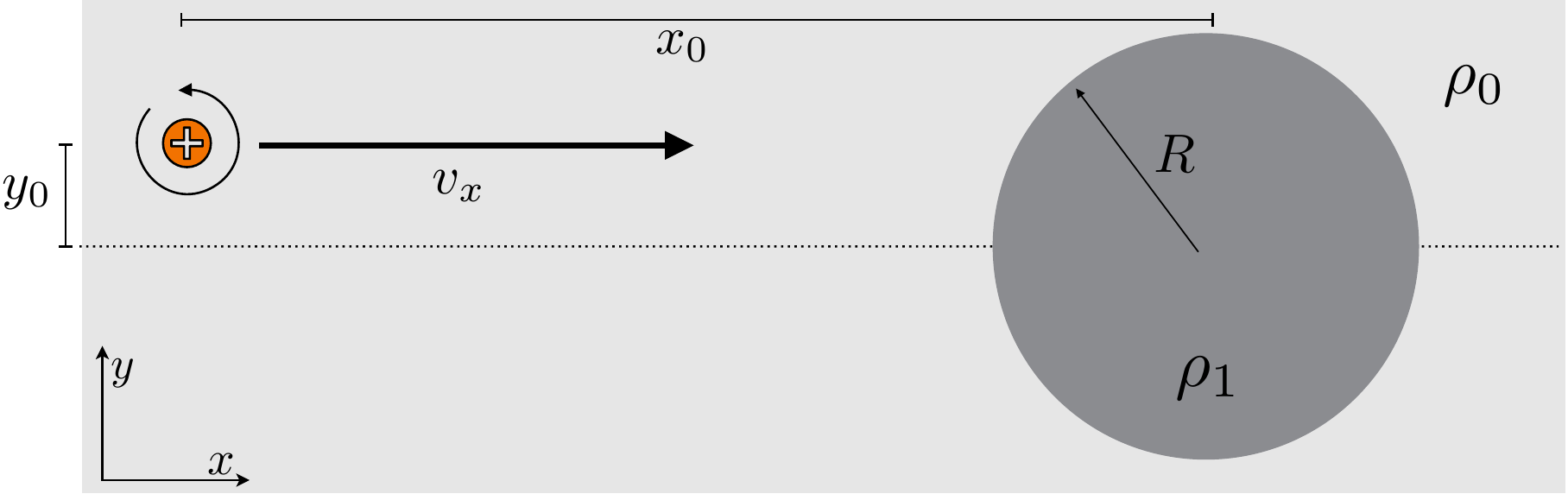}
    \caption{Vortex scattering: A superfluid with background density $\rho_0$ flows uniformly with velocity $v_x$. Embedded in the flow is a single vortex initially located at $(x_0,y_0)$.  It encounters a circular pinning potential of radius $R$ at the origin with density $\rho_1 < \rho_0$.}
    \label{fig:schematic}
\end{figure}

In this Letter, we theoretically and numerically study vortex pinning via a
 vortex ``scattering experiment" within a zero temperature superfluid, where a single vortex interacts with a circular pinning potential [Fig.~\ref{fig:schematic}].  We  develop an
analytical hydrodynamic approximation to describe the vortex
dynamics,
finding excellent agreement with Gross-Pitaevskii simulations at low
velocities. At higher velocities, we identify two mechanisms of
vortex pinning, which are clearly distinguished by the acoustic energy signals produced during the pinning process. Finally, we construct a
phase diagram of the pinning process and find that larger obstacles are
comparatively ineffective for vortex capture relative to their size.

\textit{Model}---We consider a weakly-interacting Bose gas described by the Gross-Pitaevksii equation (GPE) $i \hbar \partial_t \psi(\rr,t) = \delta H /\delta \psi^*$ with Hamiltonian
\begin{gather}
H  = \int \dd\rr \left(\frac{\hbar^2}{2m} |\nabla \psi(\rr,t)|^2 + V(\rr)|\psi(\rr,t)|^2 + \frac{g}{2} |\psi(\rr,t)|^4  \right),
\label{eqn:GPE}
\end{gather}
where $g$ characterizes the repulsive particle interactions and $m$ is the particle mass.  We assume tight confinement along the $z$-axis such that the problem effectively becomes two dimensional~\cite{rooney2011,bradley2012}. The superfluid flows along the $x$ direction with background velocity $v_x$,  past a stationary pinning potential $V(\vb{r}) = V_0(1+\tanh\qty[\qty(R-|\mathbf{r}|)/w])/2,$ with strength $V_0$, radius $R$, and boundary width $w\ll R$. A vortex is initialised at $\vb{r}_0 = (x_0,y_0)$, where $y_0$ defines the impact parameter of the scattering problem [Fig.~\ref{fig:schematic}]. 
\begin{figure}
    \centering
    \includegraphics[width=\columnwidth]{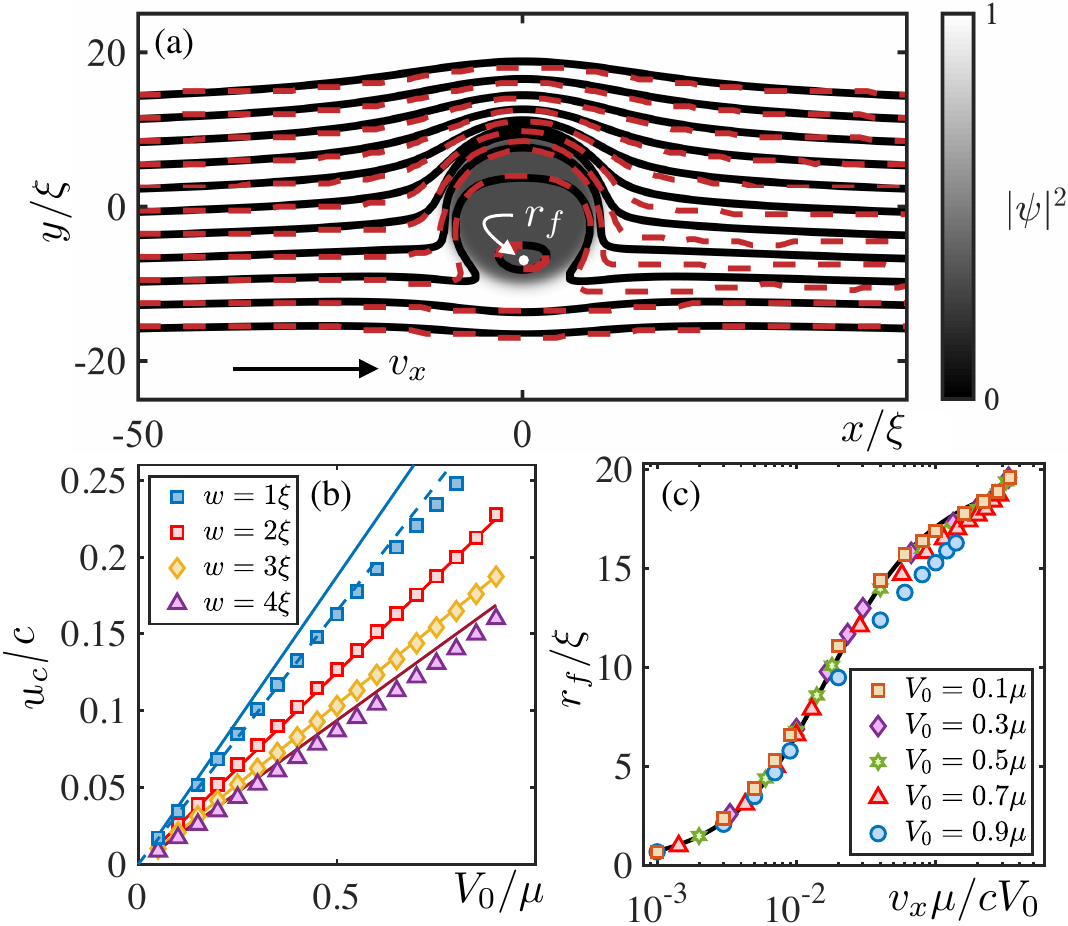}
    \caption{Hydrodynamic theory and GPE comparison. (a)  Vortex trajectories in the hydrodynamic theory (solid, black) and GPE (dashed, red) with  $v_x=0.1c$, $V_0=0.7\mu$, and $R=10\xi$.  (b) Unpinning velocity $u_c$ vs.~obstacle strength $V_0$ for boundary widths $w$; solid lines show Eq.~(\ref{eqn:uc}) with $\delta = 4.5\xi$ and $R=20\xi$. Dashed line shows the correction $w \rightarrow w_{\rm eff} = 1.2 \xi$. (c) Fixed point radius $r_f$ vs.~scaled flow velocity for $R=20\xi$ and $w= \xi$. Points: GPE solutions; line: Eq.~(\ref{eqn:rf}) with $\delta_\text{eff} = 1.5\xi$ .}
    \label{fig:trajandem}
\end{figure}

\textit{Hydrodynamic approximation}---We first consider a hydrodynamic approximation for the vortex dynamics. This approach is valid for large ($R \gg \xi$) and weak pins ($V_0 \ll \mu$), and slow velocities ($v_x \ll c$), where $\xi = \hbar/ \! \sqrt{m \mu}$ is the healing length, $\mu = n_0 g$ is the chemical potential, and $c= \sqrt{\mu/m}$ is the speed of sound.   

Under the Madelung transformation $\psi(\rr) = \sqrt{\rho/m}\, e^{i \Phi}$, the GPE can be recast to  hydrodynamic equations governing the superfluid density $\rho(\rr)$ and velocity field $\uu(\rr) = \hbar\grad \Phi /m$ with phase $\Phi(\rr)$~\cite{bradley2012}. The dynamics of a single (positive) vortex are then expressed exactly in terms of the phase and density gradients~\cite{tornkvist1997,groszek2018}
\begin{align}
\dot \rr_0 = \uu_{\Phi} + \uu_\rho \equiv   \frac{\hbar}{m} \left(\nabla \Phi|_{\rr_0} -\frac{1}{2}\vu{z}\times\grad \ln \rho |_{\rr_0} \right). \label{eqn:vortexDynamics}
\end{align}
The gradients are evaluated at the vortex position, neglecting the (singlar) self-contribution.
A steady flow with $\dot \rr_0 = \mathbf{0}$ must satisfy the mass continuity and vorticity quantization conditions
\begin{align}
\grad \cdot \mathbf{J} &= 0, & (\nabla \times \uu)_z = \Gamma \delta(\rr - \rr_0),
\label{eqn:maxwell}
\end{align}
where $\mathbf{J}(\rr) \equiv \rho(\rr) \uu(\rr)$ and $\Gamma \equiv h/m$ is the quantum of circulation. For $R \gg w \gtrsim \xi$, we may neglect the density gradients far from the  obstacle boundary, and approximate the density as a step function, $\rho(\rr) = \rho_0$ for $r > R$ and $\rho(\rr) = \rho_0(1 - V_0/\mu)$ for $r< R$. Due to the two-dimensional nature of the problem, it may be solved by a direct correspondence with electromagnetism;  defining $\mathbf{D} = -\hat{\mathbf{z}} \times  \uu$, $\vb{E} = - \hat{\mathbf{z}} \times \mathbf{J}/\rho_0$, and $\varepsilon(\rr)= \rho_0/\rho(\rr)$ with $\mathbf{D(\rr)} = \varepsilon(\rr) \mathbf{E}(\rr)$, Eqs.~(\ref{eqn:maxwell}) take the same form as Maxwell's equations in linear dielectric media. The inverse density assumes the role of the relative permittivity, and vorticity that of the free charge~\cite{browne1982}. 
The imposed flow corresponds to the displacement field for a dielectric cylinder within a uniform electric field. Thus, the imposed background flow is $\uu_{\rm im} = \hbar \nabla \Phi_{\rm im} /m$ where
\begin{equation}
\frac{\hbar}{m} \Phi_{\rm im}(r,\theta) = \begin{cases}   \tfrac{2\Upsilon v_x}{(\Upsilon + 1)} r \cos \theta, & r < R; \vspace{0.2cm}\\
v_x \left( 1 + \tfrac{\Upsilon - 1}{\Upsilon+ 1} \tfrac{R^2}{r^2} \right) r \cos \theta, & r \geq R,
 \end{cases}
 \label{imposed}
\end{equation}
where $\Upsilon = \rho_0/\rho_1$. The flow is uniform inside the obstacle and enhanced from the background value $v_x$ by $2\Upsilon/(\Upsilon +1) \geq 1$. Outside, the result is similar to the solution for an impenetrable cylinder, which is recovered as $\Upsilon \rightarrow \infty$.

The velocity due to the vortex-pin interaction may be solved via the method of images. The solution is~\cite{smythe1950static}
\begin{equation}
\frac{\hbar}{m} \nabla \Phi_v = \vb{u}_{v}(r)= -\dfrac{\Gamma}{2\pi} \left(\dfrac{\Upsilon-1}{\Upsilon+1}\right) \left(\frac{1}{d} -\dfrac{\Theta(r-R)}{r} \right) \; {\boldsymbol{\hat \mathbf{\hspace{-2pt}\theta}}} 
 \label{imagesflow}
\end{equation}
where $\boldsymbol{\hat\mathbf{\theta}}$ is the unit azimuthal vector, $d(r,R) = |R^2 - r^2|/r$ is the distance from the vortex to its image located at the inverse point $\bar \rr = R^2 \rr /|r|^2$ and $\Theta(x)$ is the Heaviside step function.
\begin{figure*}
    \centering
    \includegraphics[width=17.5cm]{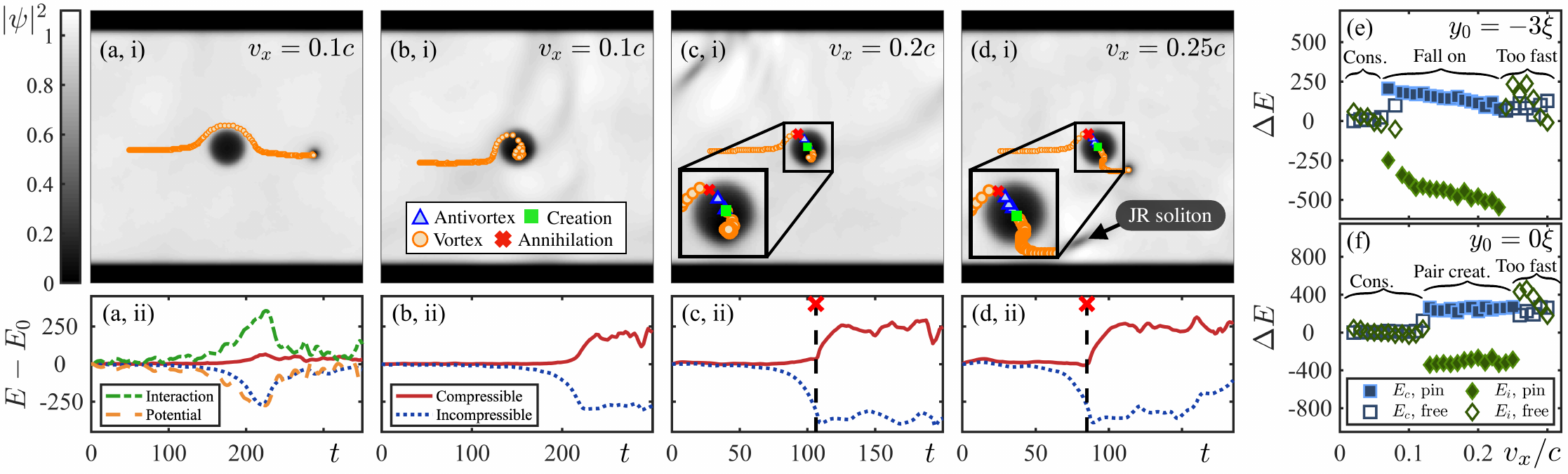}
    \caption{Regimes of vortex pinning and scattering for a pinning potential of radius $R=4\xi$ and strength $V_0=\mu$ ($w=2$): (a) Conservative, (b) Fall-on, (c) Pair creation, and (d) Too fast, (see text).  (i): Superfluid density at the end of the simulation. Markers show positions of vortices (orange circles) and antivortices (blue triangles) at equally spaced times in the dynamics. 
    For simulation movies, see~\cite{SM}. (ii): Corresponding energy exchange curves vs.~time for the simulations shown in (i). Vertical dashed lines indicate an annihilation event. For clarity, only the most relevant energy curves are displayed. Note the total energy is conserved. (e,f) Change in compressible (blue squares) and incompressible (green diamonds) kinetic energy vs.~$v_x$  for impact factors (e) $y_0=-3\xi$ and (f) $y_0=0\xi$. Solid points: vortex capture; hollow points: no capture.}
    \label{regimes}
\end{figure*}
While Eq.~(\ref{imagesflow}) diverges at $r=R$ due to neglecting the vortex core, this may be remedied by the replacement $d \rightarrow \sqrt{d^2 + \delta^2}$. The phenomenological screening parameter $\delta$ is expected to be $\delta \gtrsim \xi$, limiting the induced velocity to $|\uu_v| < c$, as required for a fluid of finite compressibility~\cite{jones1982}.  

Pinning is enabled by the existence of a fixed point $\rr_f$ where $\uu_{\rm im}(\rr_f) + \uu_v(\rr_f) + \uu_\rho(\rr_f) =\mathbf{0}$. The fixed point is located at $\rr_f=\mathbf{0}$ when $v_x =0$. As $v_x$ increases $|\rr_f|$ will approach $R$, and eventually vanish, causing the vortex to unpin. Since this occurs when $|\rr_f| \approx R$, both density and phase gradients are important. Provided $w \gtrsim \xi$, then $\uu_\rho$ may be approximated using the Thomas-Fermi solution $\rho(\rr) = \rho_0(1-V(\rr)/\mu)$. The largest density contribution, $\max[\uu_\rho(\rr)]$, occurs at $r = R + w \log(1-V_0/\mu)^{1/4} \approx R$, while $\max[\vb{u}_{v}(\rr)]$ 
occurs at exactly $r=R$. Within a logarithmic correction, we set $\dot \rr_0 = \mathbf{0}$ at $\rr_f = (R,-\pi/2)$ to obtain the critical unpinning velocity
\begin{equation}
u_c = \frac{\Gamma}{4\pi} \left(\frac{2w + \delta}{2 w \delta}\right)\frac{V_0}{\mu}.
\label{eqn:uc}
\end{equation}
Notably, $u_c$ is independent of $R$. 

Since the density and image contributions both act to counteract the imposed flow, in Eq.~(\ref{eqn:uc}) the term in parentheses may be interpreted as a reduced, \emph{effective} screening parameter, $\delta_{\rm eff} = 2w \delta/(2 w + \delta).$ Phenomenologically absorbing the density contributions into the effective screening $\delta_{\rm eff}$, we may obtain a closed form solution for the fixed point $\rr_f$ against velocity.  Balancing Eqs.~(\ref{imposed}) and~(\ref{imagesflow}) gives $\rr_f = (r_f,\theta_f)$, with $\theta_f=-\pi/2$ and
\begin{equation}
r_f = \frac{1}{2} \left\{\sqrt{4R^2  + \left(\tfrac{\Gamma V_0}{4\pi v_x \mu} \right)^2 -\delta_{\rm eff}^2 } - \sqrt{\left(\tfrac{\Gamma V_0}{4\pi v_x \mu}\right)^2 - \delta_{\rm eff}^2}\right\}.
\label{eqn:rf}
\end{equation}
Equation (\ref{eqn:rf}) bears a resemblance to the stagnation point solution in the classical Magnus effect~\cite{batchelor2000}.  Notice that  Eq.~(\ref{eqn:rf}) produces the same critical velocity as Eq.~(\ref{eqn:uc}) (real solutions vanish above $u_c$). 

The preceding (static) analysis may be applied to determine the vortex dynamics from Eq.~(\ref{eqn:vortexDynamics}) provided the velocities are small. Figure~\ref{fig:trajandem}(a) shows excellent agreement between the hydrodynamic approximation for vortex trajectories and numerical GPE solutions for an example obstacle 
(for numerical methods, see~\cite{SM}). The trajectories are asymmetric in $y$ and may be open or closed depending on the initial location of the vortex. Within the hydrodynamic approximation, the open and closed trajectories do not overlap; the vortex is thus only pinned if it is initialized on a closed trajectory inside the pin. 

Figure~\ref{fig:trajandem}(b) compares the unpinning velocity $u_c$ determined by GPE simulations~\cite{SM} to the predictions of Eq.~(\ref{eqn:uc}). We find good agreement between the two for a fixed value $\delta = 4.5\xi$. The agreement is excellent for $w/\xi\sim2$--$3$ where the assumption $\xi \lesssim w \ll R$ is valid, but  poorer for $w=\xi$~ due to a breakdown of the Thomas-Fermi approximation; however, maintaining $\delta=4.5\xi$ and replacing $w\rightarrow w_\text{eff}\approx 1.2\xi$, yields excellent agreement (equivalent to $\delta_{\rm eff} \approx 1.56 \xi$, dashed line). 

In Fig.~\ref{fig:trajandem}(c) we compare the location of the fixed point $r_f$ against the prediction Eq.~(\ref{eqn:rf}).  By scaling the background velocity with the strength of the pinning potential, the data collapse onto a single curve, with $\delta_{\rm{eff}} \approx 1.5\xi$.  The results begin to deviate for $V_0=0.9\mu$, where the assumption $V_0\ll \mu$ is no longer valid~\cite{SM}. The value $\delta_{\rm{eff}} = 1.5\xi$ obtained in Fig.~\ref{fig:trajandem}(c) is in good agreement with the best-fit observed slope of $u_c$ determined in Fig.~\ref{fig:trajandem}(b) ($\delta_{\rm eff} = 1.56\xi$).

\textit{Transition to pinning}---The hydrodynamic approach explains the conservative vortex trajectories and the mechanism for vortex unpinning. However, it cannot describe the dynamics of vortex pinning, where a vortex transitions from an open trajectory to a closed trajectory. We have therefore numerically simulated vortex scattering using the GPE over a wide range of parameters to identify how pinning occurs. To understand the dynamics of pinning, we consider the energy exchange in the system.
Equation~(\ref{eqn:GPE}) may be written as $H = E_{\rm kin} + E_{\rm pot} + E_{\rm int}$ describing the kinetic, potential, and interaction energies, respectively.  The kinetic term may be further decomposed into incompressible and compressible components~\cite{nore1997}; the incompressible part is associated with vortices, while the compressible part is due to sound waves~\cite{bradley2012}. 

We find that vortex scattering can be broadly classified into four regimes, characterized by different vortex trajectories and signatures of energy exchange.  Examples of each regime are shown in Fig.~\ref{regimes} for an obstacle with $R= 4\xi$ and $V_0 = \mu$. Figures~\ref{regimes}(a--d,i) show typical vortex trajectories, while Figs.~\ref{regimes}(a--d,ii) show the salient features of energy exchange. Note the total energy is conserved in the simulations.

\textit{1. Conservative (no pinning):} For a low background velocity $v_x = 0.1c$ and $y_0=0$, the vortex moves quasi-adiabatically around the pinning potential [Fig.~\ref{regimes}(a,i)]. The trajectories qualitatively follow the hydrodynamic curves [Fig.~\ref{fig:trajandem}(a)] and there is a near-reversible exchange between the energy components [Fig.~\ref{regimes}(a,ii)]. A small amount of energy is irreversibly lost to acoustic radiation, which slightly deflects the vortex from its hydrodynamic trajectory. This loss reduces as $v_x$ decreases;  for sufficiently low $v_x$, no pinning occurs regardless of initial conditions.

\textit{2. Fall-on (pinning):} For the same background velocity ($v_x = 0.1c$) but a sufficiently negative impact parameter ($y_0=-4\xi$), the vortex falls onto the pinning potential and enters a closed orbit [\fig{regimes}(b,i)]. The fall-on is accompanied by an 
exchange between incompressible (vortex) and compressible (sound) kinetic energies [Fig.~\ref{regimes}(b,ii)]. The occurrence of this regime, which strongly depends on the value of $y_0$, can be understood from the trajectories of the hydrodynamic theory [Fig.~\ref{fig:trajandem}(a)]. For $-R \lesssim y_0 < 0$, the trajectories exhibit high curvature implying a large acceleration. Similar to point charges, the energy radiated as sound by a line vortex is proportional to its acceleration~\cite{vinen2001}. These highly curved trajectories produce sufficient radiation to deflect the vortex onto a pinned orbit. 

\textit{3. Pair creation (pinning):} At larger velocities pinning may occur by a qualitatively different process. As shown in Fig.~\ref{regimes}(c,i) for $v_x=0.2c$ and $y_0 =0$, the incident vortex can induce the formation of a vortex-antivortex pair within the pinning potential as it approaches. The incident vortex then annihilates with the spawned antivortex, marked by a larger and more rapid burst of sound energy [Fig.~\ref{regimes}(c,ii)] than the fall-on regime. The pinned vortex is not the original vortex, but the (same-sign) vortex which spawned inside the pinning potential [\fig{regimes}(c,i)].


\textit{4. Too fast (no pinning):} For sufficiently large $v_x$ (but $v_x < u_\text{crit}$), pinning is no longer possible. The process of pair creation still occurs, but insufficient energy is radiated for the remaining vortex to be pinned and it instead escapes into the bulk [\fig{regimes}(d,i)], recovering incompressible energy [Fig.~\ref{regimes}(d,ii)]. The acoustic pulse generated by the annihilation does not always disperse into ordinary phonons; in some cases, it creates a localized, nonlinear sound pulse, known as a  Jones-Robert soliton~\cite{jones1982} [\fig{regimes}(d,ii)].

The fall-on and pair creation mechanisms are quantitatively distinct in terms of their acoustic emission signatures. In Fig.~\ref{regimes}, we show the change in the compressible and incompressible kinetic energies against $v_x$ for the impact parameters (e) $y_0 = -3\xi$ and (f) $y_0 = 0 \xi$, which exhibit the fall-on and pair creation pinning regimes, respectively. In the fall-on regime [Fig.~\ref{regimes}(e)], increasing $v_x$ reduces the curvature of the vortex trajectories [cf. Fig.~\ref{fig:trajandem}(a)] as the imposed flow dominates over the velocities induced by the pin. The amount of acoustic radiation therefore \emph{decreases} with increasing velocity, until the amount emitted is no longer sufficient to deflect the vortex onto a pinned trajectory. By contrast, the threshold velocity for pair-creation is larger than for fall-on, and the acoustic energy production is nearly constant in $v_x$ [Fig.~\ref{regimes}(f)].  At the transition to the `too fast' regime in both instances, the vortex fixed point exists, but within the boundary layer. Pinning thus becomes increasingly difficult as phonons can easily knock the vortex into the bulk.

\begin{figure}
    \centering
    \includegraphics[width=\columnwidth]{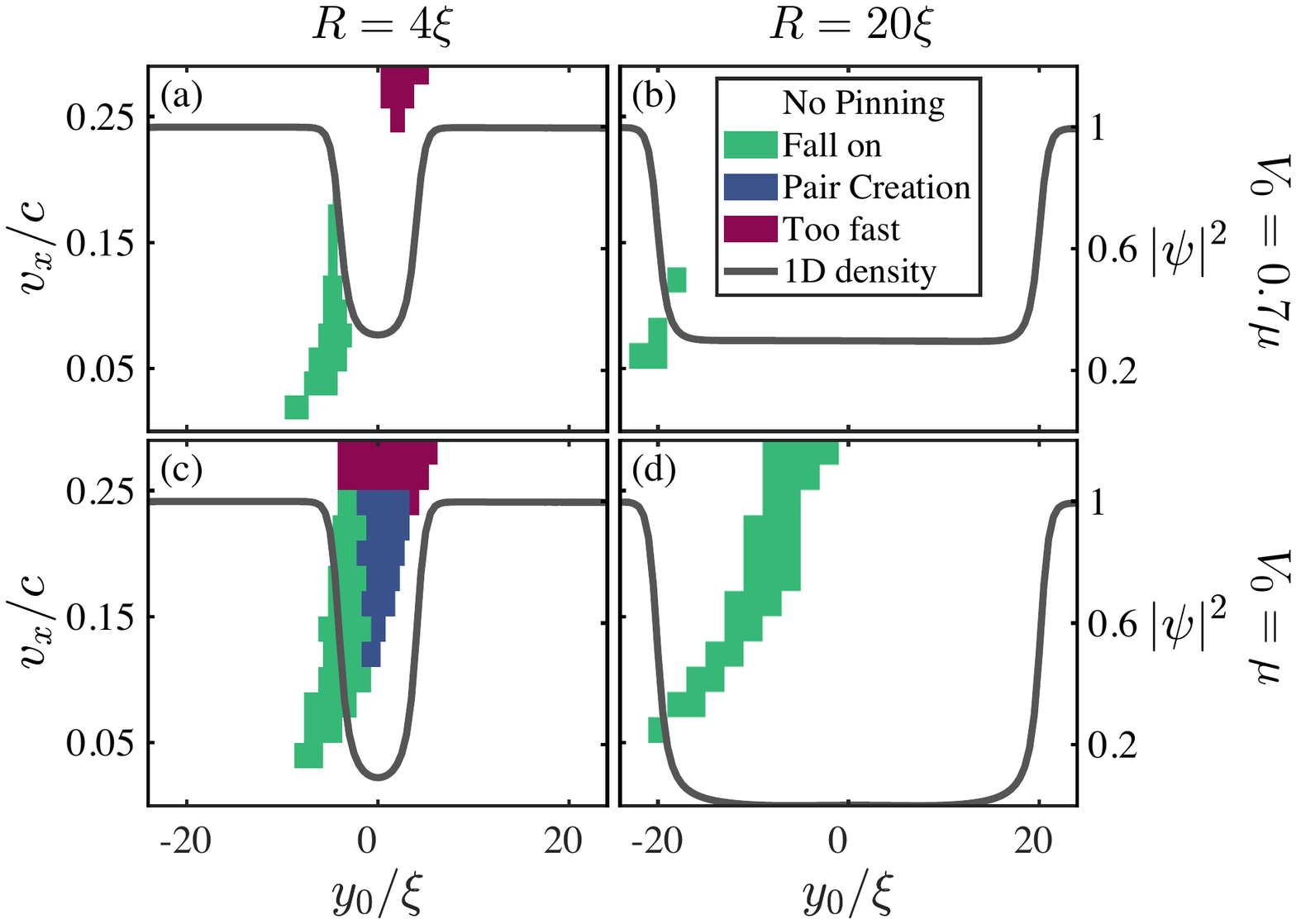}
    \caption{Vortex pinning phase diagrams as a function of impact factor $y_0$ and background superfluid velocity $v_x$ for pinning potentials with $R/\xi = \{4,20\}$ (columns) and $V_0/\mu = \{0.7,1\}$ (rows), respectively. The shaded regions represent pinning regimes (left $y$-axis) and the grey lines represent the initial superfluid density along $x=0$ at the location of the pinning potential (right $y$-axis).}
    \label{phase}
\end{figure}

Finally, in addition to $v_x$ and $y_0$, the pinning dynamics also strongly depend on the pinning potential parameters. In Fig.~\ref{phase} we plot the pinning phase diagram as a function of $v_x$ and $y_0$ for the potential radii $R=\{4,20\}\xi$ and strengths $V_0 = \{0.7,1.0\}\mu$.
The phase diagrams show that only a finite range of velocities lead to pinning. At low $v_x$, a vortex can become pinned even when the impact parameter exceeds the pinning potential radius, i.e., $y_0< -R$ [\fig{phase}(a)]. The $y_0<0$ region of parameter space is dominated by the `fall-on' regime due to the large curvature of trajectories. Conversely, vortices approaching from $y_0>0$ may only become pinned via the pair creation mechanism as described above.  This distinction can be understood in terms of a destructive ($y_0<0$) and constructive ($y_0\geq0$) vector sum of the vortex velocity field and the imposed flow field (see~\cite{SM}).

For $V_0<\mu$, increasing $V_0$ significantly increases the likelihood of pinning~\footnote{For $V_0 \geq \mu$, the pinning diagram does not change significantly.}.  The reduced superfluid density inside the pinning potential lowers the energy cost for pair creation, allowing it to occur over a wider range of parameters [\fig{phase}(c), cf.\ Fig.~\ref{phase}(a)].  Further, we find that pinning potentials of smaller radii have a larger capture area relative to their size  --- for large radii, pinning does not occur at any $V_0$ for $y_0>0$ [\fig{phase}(b,d)]. While large radii potentials provide a deep energy minimum~\cite{SM}, the pinned vortex states are less accessible dynamically; the larger radius vortex trajectories have a smaller acceleration and hence radiate less energy as sound, meaning that a relatively smaller region of phase space leads to pinning.

\textit{Conclusions and outlook}---We have considered the scattering of a quantised vortex off a repulsive pinning potential in a superfluid thin film. The hydrodynamic approximation describes the low velocity trajectories, as well as the location of the vortex fixed point and the unpinning velocity. Our predictions could be readily tested in ultracold atom experiments employing configurable optical potentials~\cite{gauthier2016}.  Radially inhomogeneous pins, which may be considered by analogy with antennae and waveguides in the dielectric analogy~\cite{vigants1962,yamashita1981}, could give different density and phase contributions that may lead to more effective pinning. 

The GPE simulations identified two distinct pinning mechanisms, each marked by a characteristic emission of sound energy. The pair creation regime resembles virtual particle interactions in high energy physics, suggesting another potentially useful analogy. In the pinning phase diagram we found that strong yet small radii potentials have a larger capture area compared to their size, being able to capture vortices at both positive and negative impact parameters. This may have important implications for devices utilising superfluid helium thin films~\cite{sachkou2019,varga2020},  where $\xi \sim 1$\AA. Our results suggest that atomic defects may be superior pins to fabricated, microscale defects; an array of small pins may prove more effective than a large pin occupying the same area. 
Like in superconducting devices, superfluid vortex pinning will likely play an important role in capturing free vortices and suppressing the breakdown of lossless flow in devices leveraging superfluidity.

\begin{acknowledgments}  
We thank Arghavan Safavi-Naini and Mark Baker for useful discussions, and Andrew Groszek,  Chris Baker, and Lewis Williamson for a critical reading of the manuscript. This research was supported by the Australian Research Council Centre of Excellence in Future Low-Energy Electronics Technologies (Project No. CE170100039) and funded by the Australian government.
\end{acknowledgments}  

\bibliography{references.bib}

\clearpage 

\onecolumngrid
\vspace{\columnsep}
\begin{center}
\textbf{\large Supplemental Material for:\\
Dynamical mechanisms of vortex pinning in superfluid thin films}
\end{center}
\vspace{\columnsep}
\twocolumngrid

\setcounter{equation}{0}
\setcounter{figure}{0}
\setcounter{table}{0}
\setcounter{page}{1}
\makeatletter
\renewcommand{\theequation}{S\arabic{equation}}
\renewcommand{\thefigure}{S\arabic{figure}}

\section*{Numerical methods}
Here we provide the full details of the numerical methods used to solve  the Gross-Pitaevskii equation (GPE) to find its stationary states and to solve for the vortex scattering dynamics. 

We consider a superfluid that is tightly confined in the $z$-direction such that the resulting system is quasi two-dimensional. In the $x$-$y$ plane the superfluid is confined to a channel, which allows it to flow in the $x$-direction under  periodic boundary conditions.  We assume the superfluid is weakly interacting, and hence describe its dynamics with the two-dimensional Gross-Pitaevskii equation with Hamiltonian
\begin{align}
    H = \int \text{d}\vb{r}\ \psi^*\qty(-\frac{\hbar^2}{2m}\grad^2+V(\rr)+\frac{g_2}{2}|\psi|^2 )\psi, \label{GPEhamil}
\end{align}
where $\psi\equiv\psi(\vb{r})$ is the complex superfluid order parameter,  $m$ is the mass of the constituent particles, $V(\vb{r})$ is the external potential and $\mu$ is the chemical potential. The parameter $g_2$ is the effective 2D interaction strength; for the common case of e.g., an atomic gas subject to harmonic confinement in $z$ with frequency $\omega_z$, $g_2=\sqrt{8\pi}\hbar^2a_s/ml_z$ where $a_s$ is the $s$-wave scattering length and $l_z = \sqrt{\hbar/m\omega_z}$ is the harmonic oscillator length in the $z$-direction. The GPE assumes the $s$-wave scattering length is much smaller than the inter-particle distance, i.e., $na_s^2 \ll 1$, where $n$ is the two-dimensional number density.

We consider an approximately uniform circular pinning potential of strength $V_0$. The entire potential takes the form
\begin{multline}
    V(\vb{r}) = V_1\qty(1+\tanh\qty[\qty(y^2-\frac{w_1^2}{4})]) \\ + \frac{V_0}{2}\qty(1+\tanh\qty[\qty(R-r)/w]),\label{potential}
\end{multline}
where the first term describes the channel (constricting the flow along the $x$-direction), and the second term describes the circular uniform pinning potential of radius $R$, and $r \equiv |\vb{r}|$. The parameter $w$ sets the steepness of the pinning potential and $w_1$ is the width of the channel. In all simulations we ensure $w_1\geq6R$ so the effect of the channel walls upon the vortex dynamics is minimal.

Due to the single-valued nature of $\psi(\rr)$ and periodic boundary conditions $\psi(x,-L_x/2)~=~\psi(x,L_xy/2)$, stationary states of the superfluid are constrained to flow along $x$ with quantized velocity $v_x = 2\pi n \hbar/ m L_x $, where $n\in\mathbb{Z}$ and $L_x$ is the length of the channel. However the GPE is Galilean invariant and we may therefore boost into a frame moving at velocity $\vb{v} = (v_f,0)$, effectively dragging the potential through the superfluid at the same velocity. This creates an effective superfluid flow past the pinning potential (which is stationary in the moving frame) and allows for arbitrary control of the effective background velocity. 

The energy in the boosted frame $H\rightarrow H-\vb{p}\cdot\vb{v}$ is 
\begin{align}
    H = \int \text{d}\vb{r}\ \psi^*\qty(-\frac{\hbar^2}{2m}\grad^2+V(\rr)+\frac{g_2}{2}|\psi|^2+i\hbar v_f\partial_x)\psi, \label{movingHamil}
\end{align}
with corresponding equation of motion
\begin{align}
i\hbar\partial_t\psi = \qty[-\frac{\hbar^2}{2m}\grad^2+V(\vb{r})+g_2|\psi|^2+i\hbar v_f\partial_{x}-\mu]\psi,\label{GPE1}
\end{align}
Note that, due to the periodic boundary conditions, the term involving $v_f$ cannot be absorbed into a phase factor for arbitrary $v_f$, as is commonly done in a free-space geometry~\cite{coleman2015}. The term involving $\mu$ serves to remove the trivial global phase evolution of the stationary states, which evolve as $\psi(\rr,t) = \psi(\rr)e^{-i\mu t /\hbar}$.

\begin{figure}
    \centering
    \includegraphics[width=7cm]{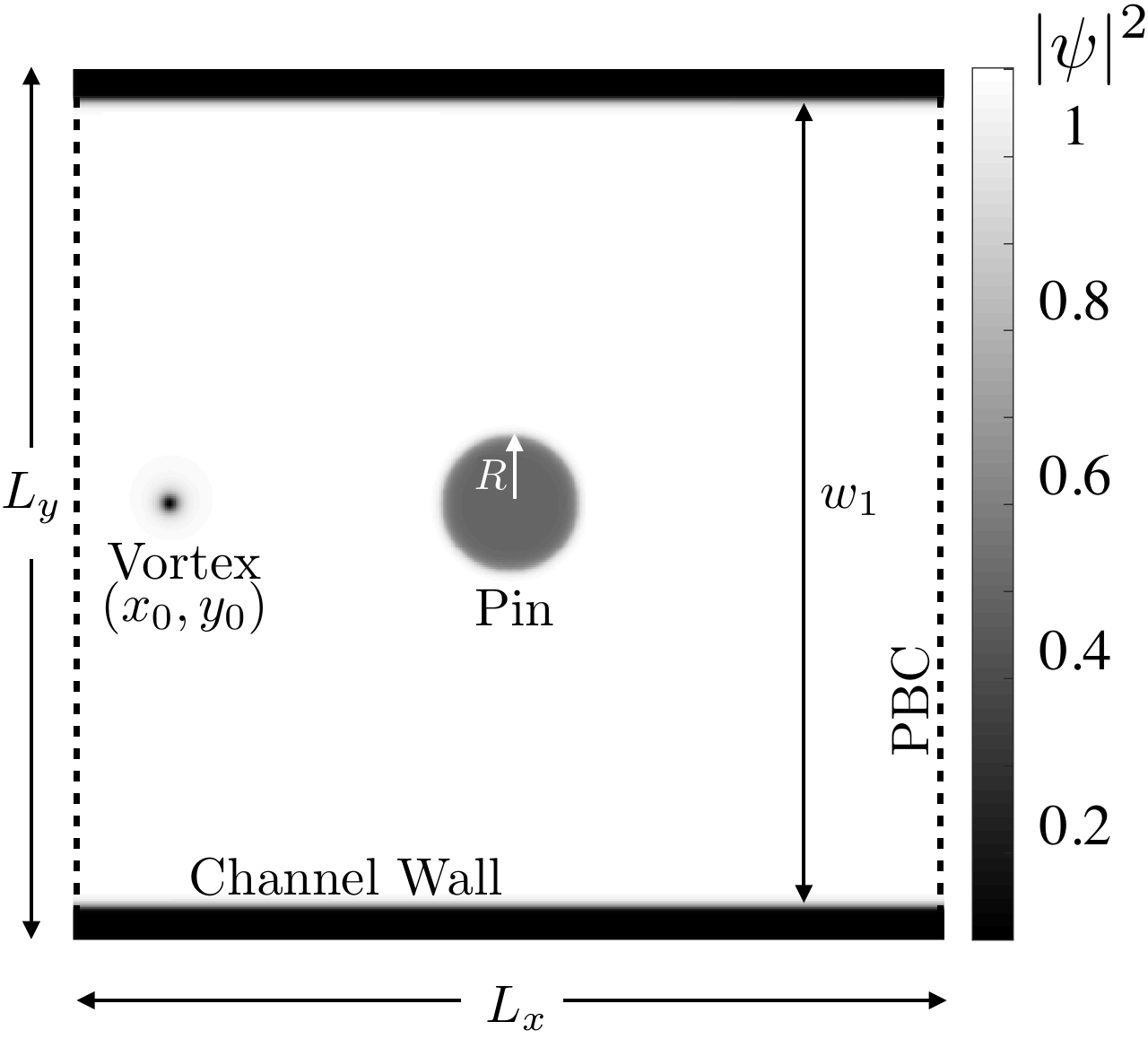}
    \caption{Example of the initial superfluid density for GPE simulations. Dashed lines represent periodic boundary conditions (PBC).}
    \label{fig:schematicSupp}
\end{figure}

We imprint a single vortex at $\vb{r}_0=(x_0,y_0)$ using the phase ansatz $\theta=\kappa\atan[(y-y_0)/(x-x_0)]$, and initialise the wave function using the Thomas-Fermi approximation
\begin{align}
    \psi(\rr) = \sqrt{\frac{\mu-V(\vb{r})}{g_2}} e^{i\pi y/L_x}e^{i\theta}. \label{TFregime}
\end{align}
The first exponential in \eq{TFregime} ensures the periodicity of the wave function due to the presence of a single vortex, and the second describes the vortex phase. We numerically pin the vortex at $\vb{r}_0=(x_0,y_0)$, and integrate \eq{GPE1} in imaginary time with $v_f=0$ to solve for the initial ground state, an example of which is shown in Fig.~\ref{fig:schematicSupp}.

We introduce the velocity of the vortex relative to the pinning potential, $v_x = v_f + \pi/L_x$, where the $\pi/L_x$ term 
accounts for the self-induced velocity of the vortex due to the geometry of the system~\cite{guenther2017}. 

\begin{figure}
    \centering
    \includegraphics[width =0.33\textwidth]{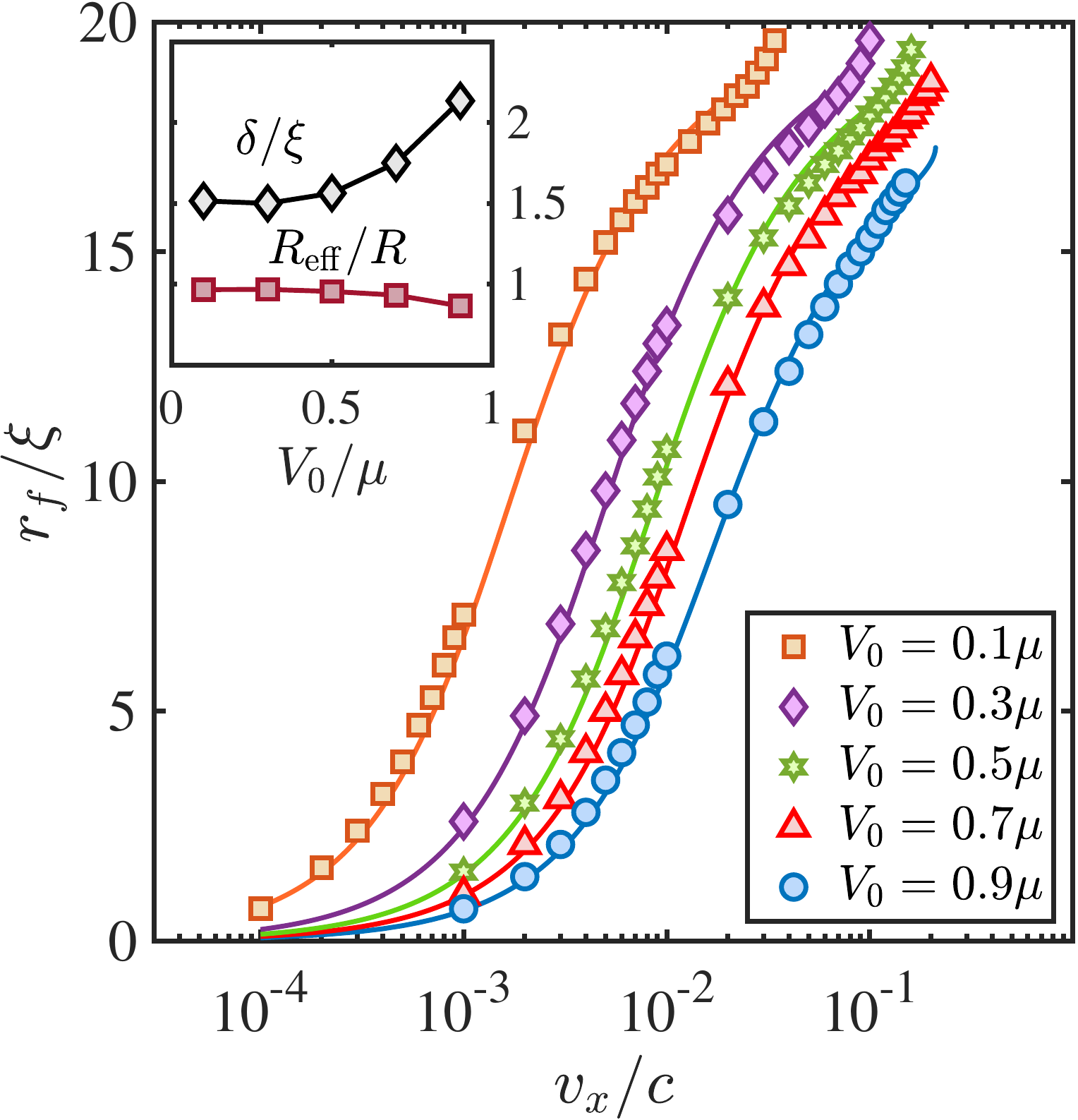}
    \caption{{Vortex fixed point radius $r_f$ as a function of flow velocity $v_x/c$ for obstacles with $R/\xi=20$ and $w/\xi=1$. The points are from stationary solutions to the GPE for a variety of obstacle strengths $V_0$, where the solid lines are fits of} Eq.~(6) in the main text to the data points with fitting parameters $\delta$ and $R_\text{eff}$. These parameters are shown in the insets. {Improved fits can be found by increasing $\delta_{\rm eff}$ and decreasing $R_{\mathrm{eff}}$ for larger values of $V_0/\mu$ as compared to the universal fit in Fig.~2(c) of the main text.}
    }
    \label{fig:new_rp}
\end{figure}

\subsubsection{Dynamics}
For simulations it is convenient to work in natural units by introducing the healing length $\xi$, speed of sound $c$, time unit $\tau\equiv\xi/c$, and chemical potential $\mu$ as the units of length, velocity, time, and energy respectively. In these units \eq{GPE1} reduces to
\begin{align}
    i\partial_t\psi = \qty[-\frac{\grad^2}{2}+V(\vb{r})+g_2|\psi|^2+i\,\qty(v_x-\frac{\pi}{L_x})\partial_x-1]\psi.\label{compGPE}
\end{align}
The interaction parameter can further be scaled out of the equation by the replacement $\psi \rightarrow \sqrt{g_2} \psi$, leading to a maximum density of $|\psi|^2=1$ in regions where $V=0$.

We solve for vortex scattering dynamics by integrating \eq{compGPE} in real time. 
We integrate \eq{compGPE} using an adaptive fourth-fifth order Runge-Kutta technique on a grid with spacing $\Delta x=\Delta y=\xi/4$ using the software package XMDS2 \cite{dennis2013}.  The magnitude of the channel potential is chosen to be $V_1 = 10\mu$, whilst the steepness of the channel potential is  chosen to be $w_1 = 1$. Unless otherwise specified, we set $w = 2$. The superfluid background velocity is linearly increased from zero with a finite acceleration $a=10^{-3}\, c^2/\xi$ for all simulations.

\subsubsection{Steady state}
The position of the vortex fixed point as a function of $v_x$ is obtained by finding the stationary solutions of the Gross-Pitaevskii equation in the translating frame [\eq{compGPE}]  using Newton-Raphson iteration with second-order finite differences.
The Newton solver requires a finer grid to ensure convergence compared to the dynamical GPE simulations. We solve the GPE on a grid of length $L_x=L_y=10R$ with spacing $\Delta x=\Delta_y=\xi/6$. The width of the channel is set to $w=8R$. Subsequently, to ensure the stability of the stationary solutions, we add random Gaussian noise to each grid point and evolve the noisy solution under \eq{compGPE}. The unpinning velocity $u_c$ is determined by the largest velocity for which a stationary solution can be found.

\section*{Fixed Point Radius}
{In Fig.~2(c) of the main text we showed that the fixed point vortex radius $r_f$  was a universal function of the background flow velocity $v_x$ when the latter quantity is scaled with the relative potential depth $V_0/\mu$. In \fig{fig:new_rp} we show the behaviour of $r_f$ without the scaling of $v_x$, but with separate fits of $\delta_{\rm eff}$ and $R_{\textrm{eff}}$ in Eq.~(6) for each $V_0$.  The inset of \fig{fig:new_rp} shows how these parameters change for different pinning depths. For $V_0\leq0.5\mu$ we find $R_{\mathrm{eff}} \approx R-w$ (with $w=\xi$) and $\delta_{\rm eff} = 1.5$.
For $V_0 > 0.5 \mu$ the hydrodynamic approximation $V_0 \ll \mu$ becomes less accurate, yet Eq.~(6) of the main text  still provides a good fit to the GPE data.  As physics beyond the hydrodynamic approximation becomes more important, we find that $R_{\mathrm{eff}}$ decreases and $\delta_{\rm eff}$ increases somewhat.}

\begin{figure}
    \centering
    \includegraphics[width=7cm]{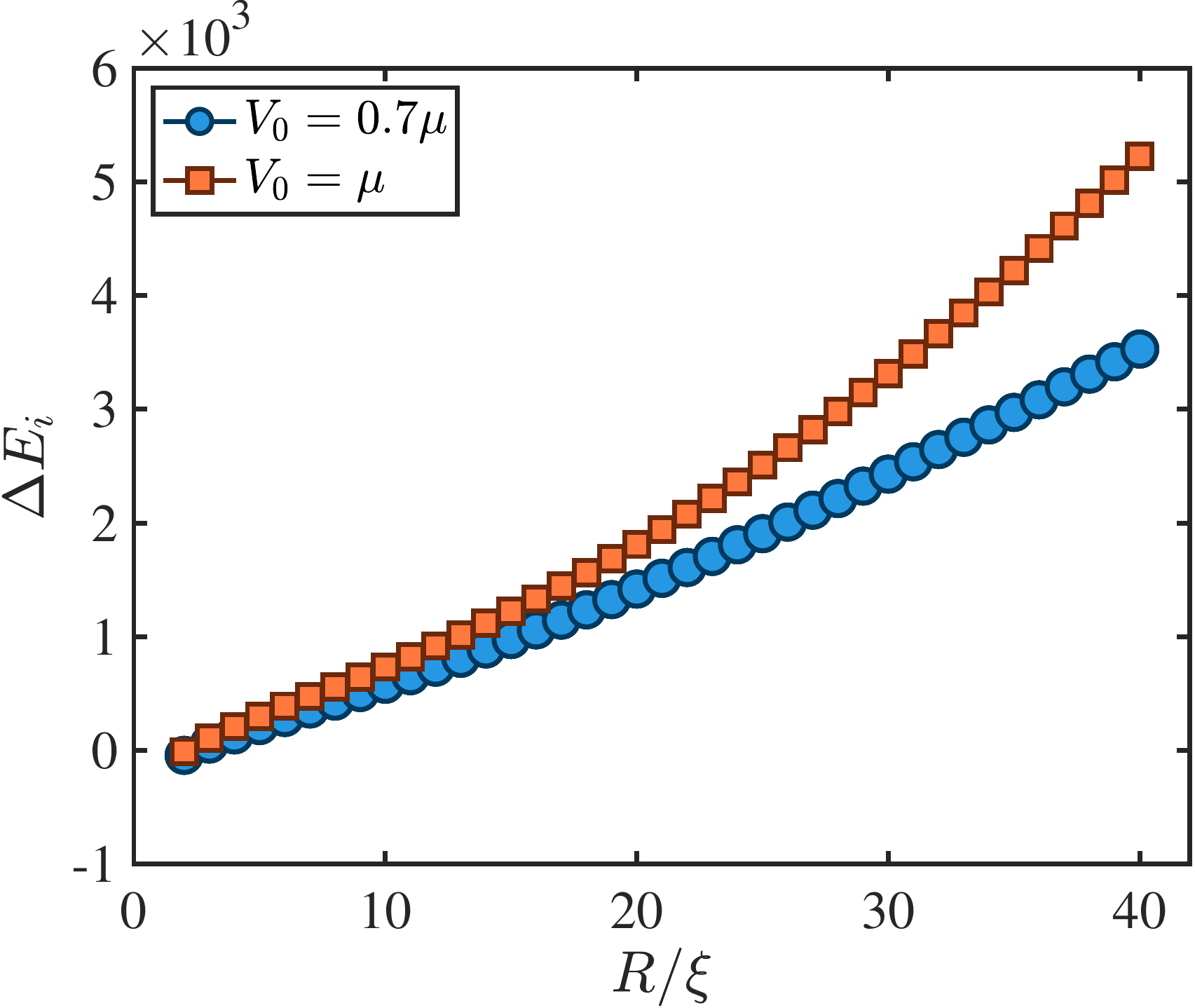}
    \caption{Difference in incompressible energy of a free vortex infinitely far away from a pinning potential and a vortex at the {fixed point} within the d{pinning potential}. {Blue circles: $V_0 = 0.7\mu$, orange squares: $V_0 = \mu$.}
   }
    \label{fig:energy}
\end{figure}

\begin{figure*}
    \centering
    \includegraphics[width = 0.9\textwidth]{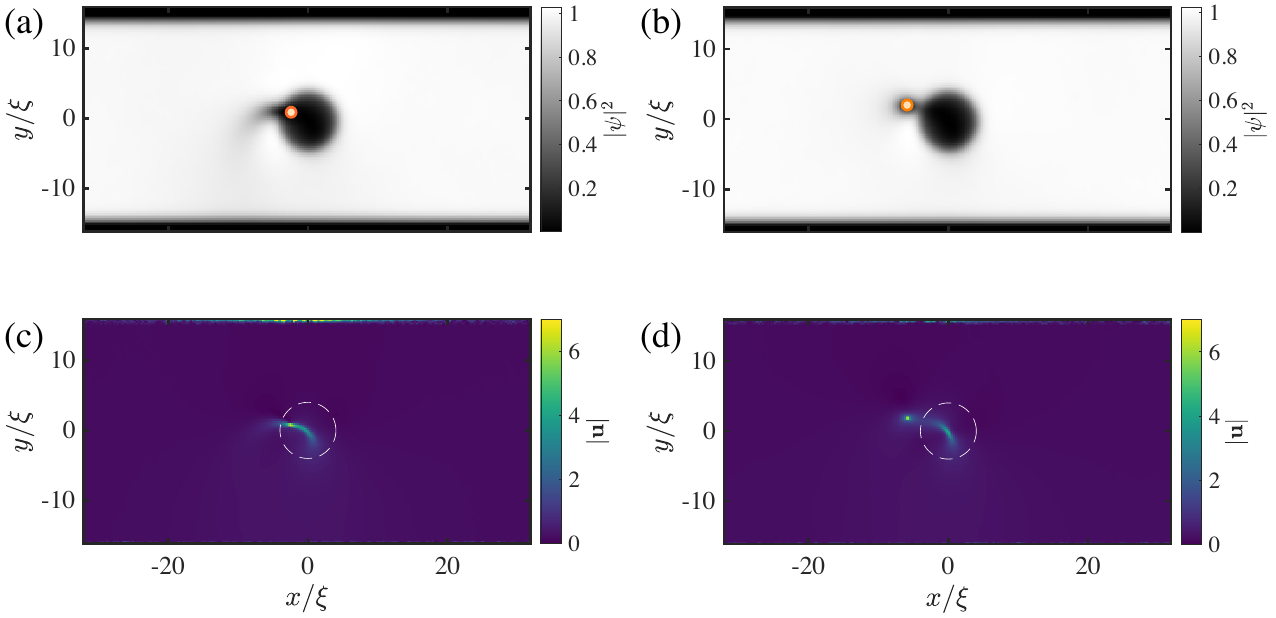}
    \caption{{Snapshots of the superfluid density (a,b) and velocity (c,d) for GPE simulations of vortex scattering in the fall-on ($v_x = 0.004, y_0 = 3$) regime  (a,c) and pair creation ($v_x = 0.008, y_0 = 0$) regime (b,d).  The plots are at a point in time where the velocity  at the centre of the pin exceeds a threshold value of $u_{\rm th }$ = 3c. The orange circle indicates the location of the incoming positive vortex.  Obstacle parameters are  $R/\xi=4$, $V_0/\mu = 0.9$ and $w/\xi = 1$.} 
    }
    \label{fig:fallon_vs_pc}
\end{figure*}

\begin{figure*}
    \centering
    \includegraphics[width = 0.9\textwidth]{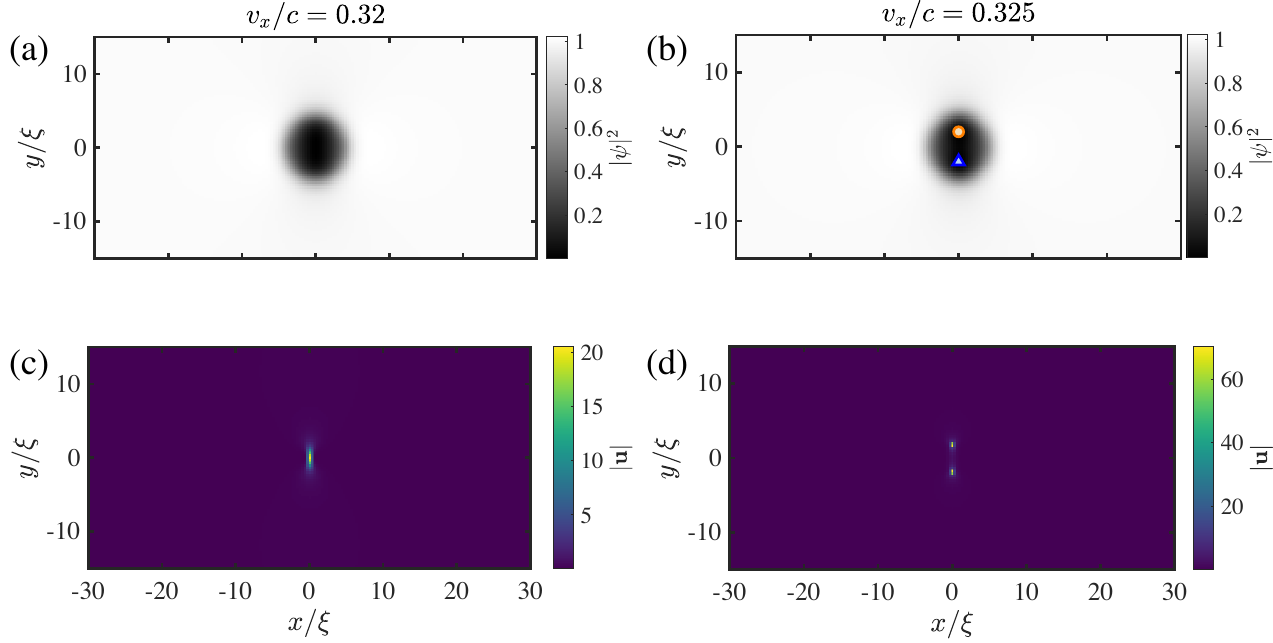}
    \caption{{Superfluid density (a,b) and velocity (c,d) of stationary solutions to the GPE in a frame moving with $v_x/c = 0.32$ (a,c) and  $v_x/c = 0.325$ (b,d) for an obstacle with $R/\xi=4$, $V_0/\mu = 0.9$ and $w/\xi = 1$.  The orange circle (blue triangle) indicates the location of a positive (negative) vortex.}}
    \label{fig:stationaryStates}
\end{figure*}

\section*{Energy of Pinned Vortex}
To expand upon the vortex pinning results presented in Fig.~4 of the main text, in \fig{fig:energy} we plot the difference in the incompressible energy between a free vortex infinitely far away from a pinning potential and a pinned vortex at the energy minima as a function of potential radius $R$. 
This difference is the amount of incompressible energy that needs to be irreversibly lost via radiation or pair annihilation in order for an initially free vortex to become pinned as per the results in Fig.~3 of the main text.  There are two curves for potential heights of $V_0 = 0.7\mu$ and $\mu$, both within a superflow of velocity $v_x = 0.1c$.


The results show that as the radius of the pinning potential increases, the amount of energy that must be lost to sound energy increases. In this pure superfluid system, there are only a finite number of mechanisms 
{for this energy loss to occur, and there is an upper limit to the energy that can be lost in a single scattering event.}
Therefore as the energy of the bound state decreases for increasing radii, it becomes increasingly difficult for a vortex to become pinned as more energy must be lost by the vortex.

\section*{Vortex pair creation mechanism}

{The mechanism for vortex pair creation in the pinning potential can be understood by considering the vector sum of the linear velocity field due to the externally imposed flow $v_x$, and the radial velocity field of the approaching vortex, $|v_r| \propto 1/r$.}
Empirically, we find that pair creation {only occurs for vortex trajectories for which $y > 0$, and it} always occurs at the centre of the pin. When $y<0$ the velocity field of the approaching positive vortex {decreases the total speed of the  flow at the pin centre. {However for $y > 0$, the approaching vortex causes the total speed of the flow at the centre of the pin to increase.  When the vortex is close enough to the pin, and the potential strength $V_0$ is sufficiently large, this can lead to a threshold velocity for vortex nucleation to be exceeded.}

Examples demonstrating how this arises are shown in Figs.~\ref{fig:fallon_vs_pc} and~\ref{fig:stationaryStates}. In Fig.~\ref{fig:fallon_vs_pc} we show an instant in time where the velocity at the centre of the pin exceeds a specified threshold (we chose $u_{\rm{th}} > 3c$). For the fall-on {regime} [Fig.~\ref{fig:fallon_vs_pc}(a)], the vortex is already well within the pin, and is releasing a burst of sound. In the pair creation case, the threshold is exceeded when the vortex is still well outside the pin. Complementary to this figure, Fig.~\ref{fig:stationaryStates} shows two stationary solutions for the same obstacle in a frame moving at constant velocity $v_x$ (here no vortex is present outside {the pinning potential}). At  $v_x/c\ = 0.32 $ [Fig.~\ref{fig:stationaryStates}(a)], the velocity field inside the pin resembles that shown in Fig.~\ref{fig:fallon_vs_pc}(d) (both resemble the velocity field of a Jones Roberts soliton~\cite{jones1982}). At slightly higher velocity $v_x/c = 0.325$, the soliton turns into  a vortex dipole.  In Fig.~\ref{fig:fallon_vs_pc}(b) the velocity structure appears at an angle, consistent with the idea of the vector sum of the velocity fields. {In the vortex scattering scenario, the approaching vortex causes an increase in the velocity of the superfluid in the pin, and induces a transition to the vortex dipole state.  The approaching vortex then interacts with this dipole, leading to the subsequent annihilation of the incoming vortex.}

\section*{Supplemental Movies}

Supplemental movies S1--S4 show examples of the vortex pinning and scattering dynamics as presented in Fig.~3(a--d) of the main text.  They correspond to the ``conservative", ``fall-on", ``pair creation", and ``too fast" regimes respectively.  The simulation parameters are provided in the caption of Fig.~3 of the main text, as well as on the title slides.

Supplemental Movie S5 shows examples of the stationary solutions of the the GPE with a vortex trapped on the pinning potential as the superfluid flow velocity $v_x$ is increased.  The pinning potential has a strength $V_0 = 0.5 \mu$, radius $R= 10 \xi$, and width $w=2$. The largest velocity corresponds to the last stable solution that is found for this pinning potential, and determines the value of the unpinning velocity $u_c$ as plotted in Fig.~2(b) of the main text.

\end{document}